\documentclass[twocolumn]{aastex631}

\usepackage{bm}

\usepackage{amsmath}
\usepackage{bbding}
\usepackage[mathlines]{lineno}
\usepackage{longtable}
\usepackage{hyperref} 
\shorttitle{}
\shortauthors{Xia et al.}

\begin{document}
\title{Revisiting the X-ray Variability Plane of AGNs: The Significant Role of the Photon Index}

\author[0009-0005-3916-1455]{Ruisong Xia}
\author[0000-0001-5525-0400]{Hao Liu\textsuperscript{\Envelope}}
\author[0000-0002-1935-8104]{Yongquan Xue\textsuperscript{\Envelope}}
\author[0009-0001-9170-3363]{Jialai Wang}
\author[0000-0002-1497-8371]{Guowei Ren}
\affiliation{Department of Astronomy, University of Science and Technology of China, Hefei 230026, China; liuhao1993@ustc.edu.cn, xuey@ustc.edu.cn}
\affiliation{School of Astronomy and Space Science, University of Science and Technology of China, Hefei 230026, China}

\author[0000-0002-0771-2153]{Mouyuan Sun}
\affiliation{Department of Astronomy, Xiamen University, Xiamen, Fujian 361005, China}

\author[0000-0002-1653-4969]{Shifu Zhu}
\author[0009-0003-5280-0755]{Mengqiu Huang}
\affiliation{Department of Astronomy, University of Science and Technology of China, Hefei 230026, China; liuhao1993@ustc.edu.cn, xuey@ustc.edu.cn}
\affiliation{School of Astronomy and Space Science, University of Science and Technology of China, Hefei 230026, China}

\author[0000-0003-4773-4987]{Qingwen Wu}
\affiliation{Department of Astronomy, School of Physics, Huazhong University of Science and Technology, Luoyu Road 1037, Wuhan, China}

\author{Xian-liang Lu}
\affiliation{School of Physics, Sun Yat-sen University, Guangzhou 510275, China}

\author[0000-0001-8515-7338]{Zhen-Bo Su}
\affiliation{Department of Astronomy, University of Science and Technology of China, Hefei 230026, China; liuhao1993@ustc.edu.cn, xuey@ustc.edu.cn}
\affiliation{School of Astronomy and Space Science, University of Science and Technology of China, Hefei 230026, China}

\author[0009-0005-2801-6594]{Shuying Zhou}
\affiliation{Department of Astronomy, Xiamen University, Xiamen, Fujian 361005, China}

\begin{abstract}
X-ray variability provides a powerful probe of the innermost regions of active galactic nuclei (AGNs), offering valuable insights into the accretion process and the structure of the corona. 
Previous studies have established a correlation between the X-ray variability timescale, black hole mass, and luminosity, forming the AGN X-ray variability plane.
A possible link between the X-ray spectral photon index and X-ray variability was noted in early studies but has rarely been incorporated into subsequent analyses of the variability plane. 
Moreover, the limited sample sizes in earlier works have limited the robustness and universality of the X-ray variability plane.
In this work, we compile a sample of 112 AGNs with 399 exposures from the 4XMM-DR14 catalog and constrain the correlations between X-ray variability timescale, black hole mass, luminosity, and photon index using the recently developed fitting method, BADDAT {(Baseline-Aware Dependence fitting for DAmping Timescales)}, which enables a robust exploration of an extended parameter space.
Our analysis confirms the dependence of the rest-frame variability timescale ($\tau_{\rm rest}$) on black hole mass ($M_{\rm BH}$) and further incorporates the photon index ($\Gamma$) into the variability plane, yielding a best-fit relation of $\log (\tau_{\rm rest}/{\rm s}) = 1.22\log (M_{\rm BH}/M_\odot) - 0.24\Gamma - 3.53$, which is strongly favored over the model with $M_{\rm BH}$ alone.
In contrast, the inclusion of luminosity does not produce a comparable improvement.
The correlation with $\Gamma$ likely reflects the effects of Comptonization and the geometry of the corona.
\end{abstract}


\section{INTRODUCTION}\label{sec1}
% Corona emits X-rays
Active galactic nuclei (AGNs) are powered by accretion {of matter} onto supermassive black holes, where the innermost regions of the accretion flow produce intense high-energy radiation.
A significant fraction of X-ray emission is believed to arise from a hot, optically thin corona, in which thermal photons from the accretion disk are {inverse} Compton up-scattered to X-ray energies.
X-ray variability offers a powerful probe of the inner regions of {AGNs}, providing insight into the accretion process and the structure of the corona \citep[e.g.,][]{2005MNRAS.359..345U, mchardy_Active_2006}.

% PSD
Most X-ray observations exhibit stochastic variability.
On timescales ranging from minutes to days, this variability is characterized by a power spectral density similar to red noise \citep{gonzalezmartin_xray_2012}.
In their sample of 104 AGNs, \citet{gonzalezmartin_xray_2012} found that most power spectra are well fitted by a simple power-law model with an average photon index of $2.01 \pm 0.01$, while a subset of sources, mainly Type I AGNs, is better described by a bending power-law model whose bending frequency ($\nu_{\rm br}$) appears to correlate with black hole mass ($M_{\rm BH}$) and bolometric luminosity ($L_{\rm bol}$).

% MBH-Timescale
\citet{markowitz_xray_2003} identified a positive correlation between the X-ray variability break timescale ($T_{\rm b} = 1/\nu_{\rm br}$) and black hole mass using a sample of several Seyfert 1 galaxies, with $T_{\rm b}/\rm d=M_{\rm BH}/(10^{6.5}M_\odot)$.
% MBH-Timescale-L
Using a broader sample that included both AGNs and Galactic black holes, \citet{mchardy_Active_2006} demonstrated that the break timescale can be more precisely constrained when the dependence on luminosity is taken into account, yielding 
$\log (T_{\rm b}/{\rm d}) = 2.10 \log (M_{\rm BH}/(10^6 M_\odot)) - 0.98 \log (L_{\rm bol}/(10^{44}\rm\ erg\ s^{-1})) - 2.32$.
This empirical relation, commonly referred to as the variability plane, has reinforced the view that AGNs are fundamentally scaled-up counterparts of Galactic black holes \citep{mchardy_Active_2006}.
Using an updated sample, \citet{gonzalezmartin_xray_2012} obtained the relationship $ \log (T_{\rm b}/{\rm d}) = 1.34 \log (M_{\rm BH}/(10^6 M_\odot)) - 0.24 \log (L_{\rm bol}/(10^{44}\rm\ erg\ s^{-1})) - 1.88 $, where the coefficient for the luminosity term is consistent with zero within $1\sigma$. 
More recently, \citet{2025arXiv251014529L} updated the X-ray variability plane using new bend timescale measurements, finding $ \log (t_{\rm bend}/{\rm d}) = 1.2 \log (M_{\rm BH}/(10^6 M_\odot)) - 0.15 \log (L_{\rm bol}/(10^{44}\rm\ erg\ s^{-1})) - 1.8 $, which is in good agreement with the earlier result \citep{gonzalezmartin_xray_2012}.

%Variability - Spectra
X-ray variability appears to be influenced by factors beyond the commonly considered $M_{\rm BH}$ and $L_{\rm bol}$. 
For instance, in a sample of 9 AGNs, \citet{yang_exploring_2022} found that significant changes in the X-ray energy spectrum are often accompanied by substantial variations in the power spectrum. 
Similarly, \citet{2018ApJ...858....2G} reported that incorporating the photon index $\Gamma$ can slightly improve the characterization of variability, although the effect was not statistically significant, possibly due to their limited sample size.
Looking further back, \citet{1997A&A...326L..25K} identified a significant anti-correlation between the X-ray timescale and the photon index $\Gamma$, which has not been systematically revisited in recent years.
These studies motivate a more detailed investigation of X-ray variability in relation to the photon index $\Gamma$, within the framework of the previously established variability plane.

%tau_drw and BADDAT
Optical variability studies offer approaches that can be usefully adapted to the X-ray regime.
Measuring the characteristic timescale with a Gaussian process damped random walk (DRW) model has provided a powerful approach for constraining optical variability \citep[e.g.,][]{kelly_are_2009,2010ApJ...708..927K,2010ApJ...721.1014M, 2011ApJ...727L..24D, 2013ApJ...765..106Z,2021ApJ...907...96S}. 
However, the limited baselines of light curves have long posed a challenge for such analyses, often resulting in systematic underestimation of the variability timescales \citep[e.g.,][]{2010ApJ...708..927K,2024ApJ...961....5H,2024ApJ...966....8Z, 2024ApJ...975..160R}.
Recently, a population-level framework, BADDAT {(i.e., Baseline-Aware Dependence fitting for DAmping Timescales; \citealt{2025MNRAS.544L..96X})}, has been developed to address this underestimation and to provide an unbiased constraint on the dependence of variability on AGN properties.
Although originally designed for optical variability, BADDAT can be adapted to X-ray studies in the present context.

The paper is organized as follows.
In Section~\ref{sec_data}, we describe the selection of our X-ray AGN sample and the data processing. 
In Section~\ref{sec_res}, we explore correlations among key parameters and present the results of fitting a variability plane, which are discussed in Section~\ref{sec_dis} and concluded in Section~\ref{sec_con}{, respectively.}

\section{Samples and Data}\label{sec_data}

\subsection{Initial Sample}
%Select from XMM-Newton Archive
To analyze the AGN variability, we selected sources from the 4XMM-DR14 {catalog}\footnote{https://heasarc.gsfc.nasa.gov/W3Browse/xmm-newton/xmmssc.html} containing 1,035,832 detections and 692,109 sources \citep{2020A&A...641A.136W,2020A&A...641A.137T}.
We applied a conservative selection requiring the EPIC flux in the 2.0--4.5~keV band (EP\_4\_FLUX) to exceed $1 \times 10^{-13}\ \rm erg\ cm^{-2}\ s^{-1}$, ensuring reliable photon index fitting above 2~keV. 
This yielded 37,906 detections.  
We retrieved observational information from the XMM-Newton observation {catalog}\footnote{https://heasarc.gsfc.nasa.gov/W3Browse/xmm-newton/xmmmaster.html} and restricted the sample to observations with exposures longer than 10~ks, yielding 12,483 detections. 
We then cross-matched the sources with the SIMBAD database \citep{2000A&AS..143....9W} using a $3\arcsec$ matching radius to identify optical counterparts and obtain redshift measurements and classification information.
We selected sources classified as AGNs, including those identified in SIMBAD as AGN, Seyfert, Seyfert~1, Seyfert~2, Radio Galaxy, LINER, and QSO.
We excluded Blazars, as their X-ray variability is not thought to correlate with the corona.
The resulting initial sample consists of 1,429 sources {with} 2,722 detections.

\subsection{Data Processing}
%EPPROC & GTI
All observational data files (ODFs) were processed using the Science Analysis Software (SAS, version 21.0.0).
For consistency, only the EPIC-pn camera \citep{2001A&A...365L..18S} data were used.
Calibrated EPIC event files were generated from the original observational data files using \textsc{epproc}.
Only single- and double-pixel events were included (PATTERN $\leq 4$ and FLAG $= 0$).
A filter condition of RATE {$< 0.4$} in the 10--12 keV light curve was applied to define a good time interval (GTI).
%src regions
Source regions were selected within a $40\arcsec$ radius circle centered on the source of interest.
Since some observations include multiple exposures, the total number of exposures is 2,869.
In this work, each source observed in a given exposure is counted as one exposure instance, and each exposure is analyzed as an independent light curve {and  contributes one independent data point.}

%bkg regions
For each observation, we obtained a source list from the XMM-Newton pipeline processing system (PPS).
Background regions were selected as nearby circles using \textsc{ebkgreg}, with sources within these regions subtracted according to the source list from the PPS files.
%Light curves
Background-subtracted light curves were obtained from 0.3--10 keV using \textsc{evselect} and \textsc{epiclccorr} with a binning time interval of 100 seconds.

To ensure robustness in the variability analysis, we required that the light curves exhibit significant deviations from white noise, defined as an autocorrelation function inconsistent with white noise at the 3$\sigma$ level \citep{burke_characteristic_2021}.
This criterion reduced the sample to 178 sources with 549 exposures.

% Check for Pile-up
We checked the sample for pile-up effects and re-extracted the light curves and energy spectra of affected observations using an annular source region.
The outer radius was fixed at $40\arcsec$, while the inner radius was adjusted to $2.5\arcsec$, $5\arcsec$, $7.5\arcsec$, $10\arcsec$, $12.5\arcsec$, or $15\arcsec$ to mitigate pile-up effects.
From our sample, we excluded 2 sources with 2 exposures whose source regions extended to the edge or fell into the gaps of the detector.

%Spectra
The source and background spectra were extracted using \textsc{especget}, and the binned spectra were obtained with \textsc{grppha}, applying a minimum of 30 counts per bin.
PyXspec, the Python interface to the XSPEC spectral-fitting package \citep{1996ASPC..101...17A}, is employed to model the energy spectra and derive the {absorption-corrected} \hbox{2--10~keV} X-ray luminosity and photon index characterizing the hard X-ray power-law emission from the hot corona.
Two models are considered: Model-A, \texttt{TBabs*zpowerlw}, which accounts for absorption only by the local Galactic column, and Model-B, \texttt{TBabs*zTBabs*zpowerlw}, which includes absorption by both the local Galactic column and an intrinsic column. 
The hydrogen column density in the \texttt{TBabs} component of both models is fixed to the Galactic value obtained from the HI4PI survey \citep{2016A&A...594A.116H}.
Model-B is only used when the p-value from a \textit{f}-test is less than 0.01, indicating {that} it is better to use Model-B than Model-A.
The spectra are fitted in the 2--10 keV range to mitigate the impact of soft X-ray excess and absorption.
Most of them are well-fitted ($\chi^2/\text{d.o.f} < 2$), while 13 spectra failed and have been excluded from our sample, resulting in 175 sources with 534 exposures.
{Note that incomplete modeling of these spectral components can introduce biases in individual measurements. 
However, our goal is to prioritize broad statistical relations rather than the detailed spectral characterization of individual exposures. 
Therefore, we adopt a uniform analysis pipeline for all exposures, i.e., fitting with a simple absorbed power-law model. 
In the context of our analysis, such effects may represent one of the sources contributing to the uncertainties in the subsequent BADDAT regression. 
}

\subsection{Final Sample}
\begin{figure}[t]
\centering
\includegraphics[width=0.46\textwidth]{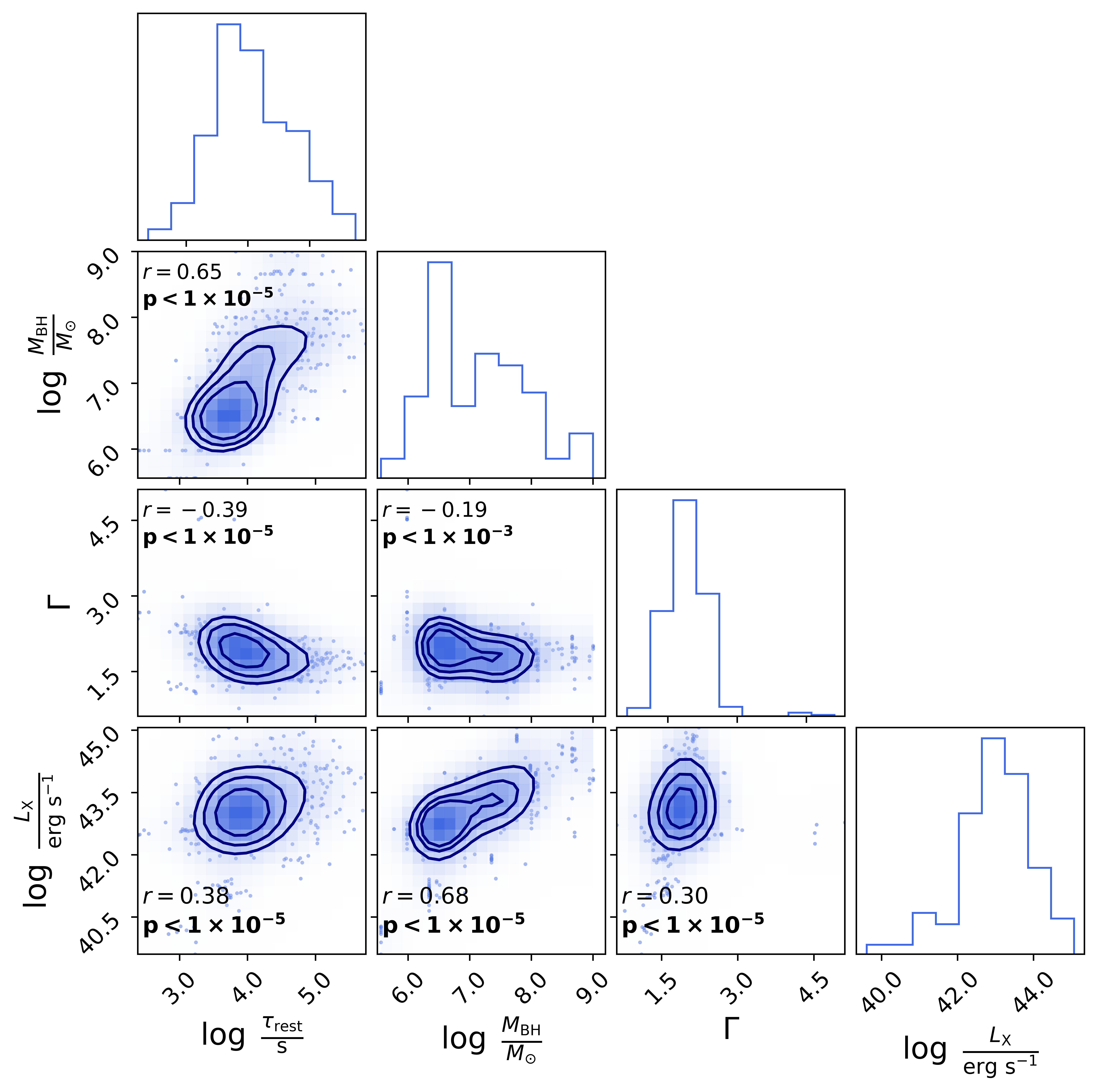}
\caption{
Correlation matrix of the timescale, black hole mass, photon index, and X-ray luminosity. 
The diagonal panels show the one-dimensional distributions of each parameter, 
while the lower off-diagonal panels display the smoothed two-dimensional density contours for each pair of variables.
The Pearson correlation coefficient ($r$) and the corresponding $p$-value {level} are indicated in each panel.
}\label{fig1}
\end{figure}

We fit the light curves with the DRW model using the maximum likelihood estimation method in \textsc{celerite} \citep{foreman-mackey_dfm_2017}.
DRW is a Gaussian process, known in the physics literature as the Ornstein-Uhlenbeck process, and is developed from the theory of Brownian motion \citep{uhlenbeck_theory_1930}. \citet{gillespie_mathematics_1996} {discussed} the mathematical details of Gaussian processes, emphasizing that a kernel function can completely determine a Gaussian process model. 
For the DRW, the kernel function, using the mathematical form from \citet{burke_characteristic_2021}, is given by:
\begin{equation}
k(t_{nm})=2\sigma^2e^{-t/\tau},
\end{equation}
where $t_{nm}=|t_n - t_m|$ is the time interval between measurements $m$ and $n${, and $\sigma$ and $\tau$ are the DRW amplitude and characteristic timescale, respectively.
As} additional white noise is considered, the kernel function becomes
\begin{equation}\label{eq3}
k(t_{nm})=2\sigma^2e^{-t/\tau}+\sigma^2_n\delta_{nm}\;,
\end{equation}
where $\sigma_n$ is the intensity of the white noise and $\delta_{nm}$ is the Kronecker delta function.

We adopted the same {model selection approach based on the Akaike Information Criterion (AIC; \citealt{1974ITAC...19..716A})} as described in {\citet{2024ApJ...975..160R} and} \citet{2025MNRAS.544L..96X}. 
The AIC is defined as 
$\text{AIC} = -2\ln(\mathcal{L}) + 2N$, 
where $\mathcal{L}$ is the maximum likelihood and $N$ is the number of free parameters. 
For each fit, the $\text{AIC}_{\rm best}$ {of the best fit} was compared with $\text{AIC}_{\rm low}$ and $\text{AIC}_{\rm hi}$, 
which correspond to the extreme cases of the fixed timescale, $\tau_{\rm out} = \text{cadence}/100$ and $\tau_{\rm out} = 100\times \text{baseline}$, respectively.
A fit was considered unreliable if either $\Delta\text{AIC}_{\rm low}$ or $\Delta\text{AIC}_{\rm hi}$ was smaller than 2, 
indicating that the best-fit model was not significantly better than these unreasonable alternatives. 
Besides, the fittings where $\tau_{\rm out} < \text{cadence}$ have been excluded due to the lack of knowledge in this {regime}.
Applying these criteria yields 156 AGNs with 451 exposures. 

We collected black hole mass measurements from the literature to construct the variability plane. 
Reliable black hole mass estimates are available for 112 sources, corresponding to 399 exposures, which are compiled in Table~\ref{tabA} in Appendix~\ref{appendixA} along with their references and constitute our final sample.
For sources without reported uncertainties, we adopted a default value of 0.1 dex in our analysis, which corresponds to the mean uncertainty of black hole mass measurements in the sample.
The distributions of the variability timescale, black hole mass, photon index, and X-ray luminosity for all exposures in the final sample are shown in the diagonal panels of Figure~\ref{fig1}.

\section{Analyses and Results}\label{sec_res}

\subsection{Correlations}

Before investigating the regression among the parameters, we first examine the pairwise correlations among them. 
Figure~\ref{fig1} presents a visual summary of the relationships among the rest-frame timescale {$\tau_{\rm rest}$ ($\tau_{\rm rest}= \tau/(1+z)$, where $z$ is the redshift)}, black hole mass $M_{\rm BH}$, X-ray photon index $\Gamma$, and X-ray 2--10~keV luminosity $L_{\rm X}$. 
The Pearson correlation coefficients and corresponding $p$-values are shown on the left side of each panel. 
All parameter pairs exhibit evident correlations ($p < 1\times10^{-2}$), with the {rest-frame timescale $\tau_{\rm rest}$} showing the strongest correlation with the black hole mass $M_{\rm BH}$. 
While previous studies mainly focused on the variability plane defined by $M_{\rm BH}$ and $L_{\rm X}$, we find that {$\tau_{\rm rest}$} is also strongly correlated with the photon index $\Gamma$, which is comparable to or even stronger than its correlation with $L_{\rm X}$.

\subsection{BADDAT Regression} \label{sec_BADDAT}
\begin{table*}[t]
    \setlength{\tabcolsep}{5mm}
    \centering
    \begin{tabular}{ccccccc}

         \hline
         \hline
         {Model}&$A$&$B$&$C$&$D$&$\sigma_\epsilon$&BIC  \\
         \hline

    1 & $1.28^{+0.11}_{-0.10}$ & --- & --- & $-4.37^{+0.65}_{-0.66}$ & $0.40^{+0.02}_{-0.02}$ & $-200.2$ \\
    2 & --- & $-0.76^{+0.10}_{-0.11}$ & --- & $5.88^{+0.24}_{-0.21}$ & $0.54^{+0.02}_{-0.02}$ & $-15.2$ \\
    3 & --- & --- & $0.45^{+0.06}_{-0.05}$ & $-14.75^{+2.25}_{-2.38}$ & $0.53^{+0.02}_{-0.02}$ & $-28.4$ \\
    $\bf 4$ & $\bf 1.22^{+0.11}_{-0.09}$ & $\bf -0.24^{+0.06}_{-0.07}$ & --- & $\bf -3.53^{+0.62}_{-0.73}$ & $\bf 0.39^{+0.02}_{-0.02}$ & $\bf -209.5$ \\
    5 & $1.54^{+0.15}_{-0.14}$ & --- & $-0.20^{+0.06}_{-0.07}$ & $2.38^{+2.32}_{-2.29}$ & $0.39^{+0.02}_{-0.02}$ & $-202.3$ \\
    6 & --- & $-0.78^{+0.09}_{-0.09}$ & $0.57^{+0.05}_{-0.05}$ & $-18.73^{+2.17}_{-2.12}$ & $0.47^{+0.02}_{-0.02}$ & $-106.9$ \\
    7 & $1.30^{+0.18}_{-0.16}$ & $-0.20^{+0.09}_{-0.08}$ & $-0.06^{+0.09}_{-0.09}$ & $-1.64^{+2.91}_{-2.82}$ & $0.39^{+0.02}_{-0.02}$ & $-201.4$ \\
    \hline
    \hline
    \end{tabular}
    \caption{Results from BADDAT regressions with different combinations of variables. 
    The fitted relation is 
    $\log(\tau_{\rm rest}/{\rm s}) = A\log(M_{\rm BH}/M_\odot) + B\Gamma + C\log(L_{\rm X}/(\rm erg\ s^{-1}))+ D$.
    Each row corresponds to a specific model configuration. 
    {Entries marked as ``--'' indicate that the corresponding variable is not included in the regression.} 
    The BIC values are used to assess the model performance, with the lowest BIC (bolded) indicating the preferred model.
    }
    \label{tab1}
\end{table*}

\begin{figure*}[t]
    \centering
    \includegraphics[width=1\linewidth]{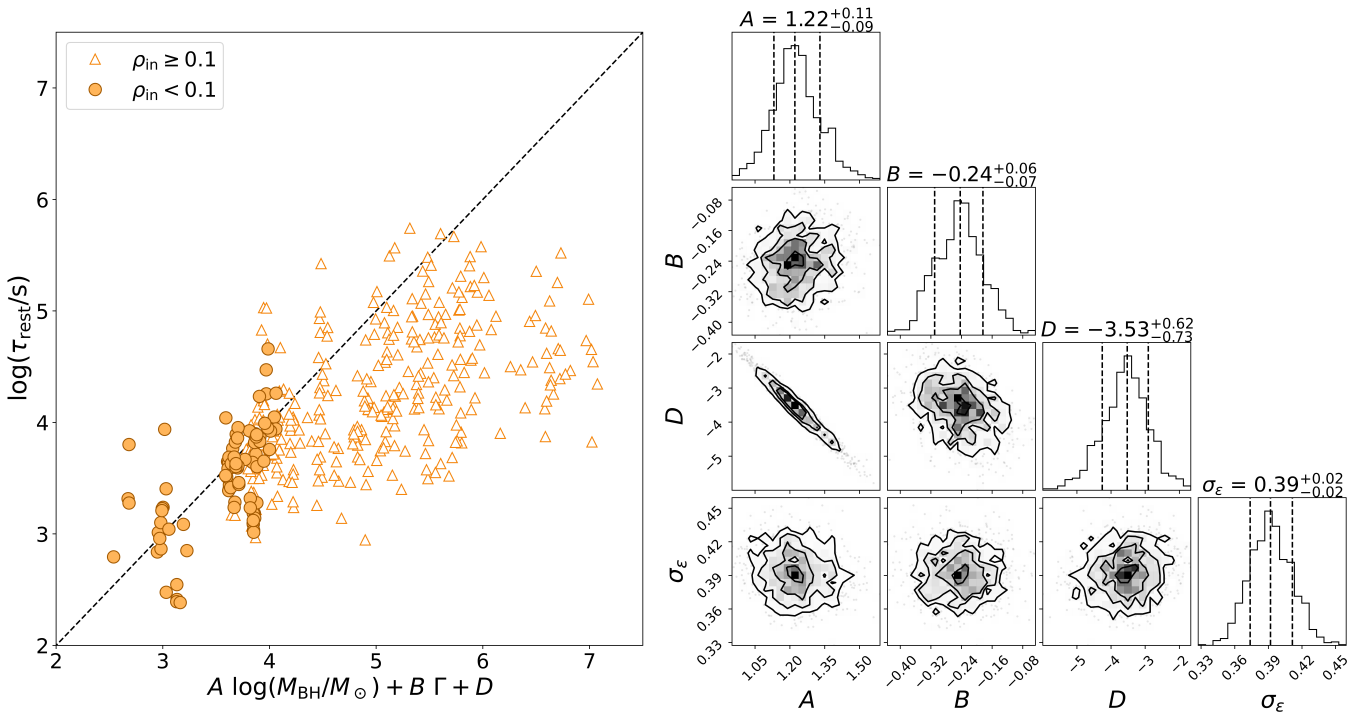}
    \caption{
    An illustration of estimating the dependence of the rest-frame variability timescale on $M_{\rm BH}$ and $\Gamma$ using BADDAT. 
    The left panel displays the rest-frame timescale as a function of {$M_{\rm BH}$ and $\Gamma$, where the black dashed line represents the relation $\log (\tau_{\rm rest}/{\rm s})=A\log(M_{\rm BH}/M_\odot)+B\Gamma+D$}.
    Data points with {modeled intrinsic timescale in the observer frame $\tau_{\rm in}=A\log(M_{\rm BH}/M_\odot)+B\Gamma+D+\log(1+z)$} shorter than 0.1$\times$baseline {(i.e., $\rho_{\rm in}<0.1$)} are shown as dots and the others {(i.e., $\rho_{\rm in}\geq0.1$)} as triangles.
    The right panel presents the probability distributions of the fitted coefficients $A$ and $B$, intercept $D$, and the scatter $\sigma_\epsilon$ and their {covariances}, obtained using our BADDAT approach with EMCEE. The 16th, 50th, and 84th percentiles of the {probability distributions} are indicated by black dashed lines, respectively.
    }
    \label{fig2}
\end{figure*}

We use the BADDAT method \citep{2025MNRAS.544L..96X} to constrain the dependence of the variability timescale on physical properties. 
Fortunately, the characteristic timescales of X-ray variability fall within the range accessible to BADDAT, 
which was originally developed for optical variability.
The implementation for X-ray variability has been updated accordingly, as described below.

We consider the likelihood function written in terms of $\rho_{{\rm out},i}$, where $\rho=\tau/\text{baseline}$:
\begin{equation}\label{eq:like}
    \ln \mathcal{L} = -\frac12\left (\sum_i\frac{[\xi(\mu_i)-\log \rho_{{\rm out},i}]^2}{s_i^2} + \ln(2\pi s_i^2)\right),
\end{equation}
where $\mu_i$ and {$\xi(\mu_i)$} are the expected value of $\log \rho_{{\rm in},i}$ and $\log \rho_{{out},i}$, respectively, corresponding to the input and output of the DRW fitting for the $i$-th {exposure}. 
The variance is given by
\begin{equation}\label{eq:sigma}
    s_{i}^2 =  [\Delta \xi(\mu_i)]^2 + \sigma_\epsilon^2 +[\xi'(\mu_i)]^2 \sum_j (k_j \Delta X_{j,i})^2,
\end{equation}
{which accounts for the combined effect of the expected uncertainty in $\xi(\mu_i)$, the noise term $\sigma_\epsilon^2$, and the propagation of uncertainties from $X_{j,i}$ to $\rho_{\rm out}$.} Details are provided in \citet{2025MNRAS.544L..96X}. 
Here we modify the treatment of the noise term $\sigma_\epsilon^2$ by moving it outside the factor $[\xi'(\mu_i)]^2$. 
This adjustment accounts for additional uncertainties beyond the intrinsic scatter of AGN variability, which may arise from measurement errors, the influence of unmodeled parameters, or deviations of the variability patterns from a DRW model.
Although the interpretation of $\sigma_\epsilon$ becomes less straightforward, it allows the algorithm greater flexibility and tolerance to such effects.

To quantify how the variability timescale $\tau$ depends on different physical properties, we performed a series of regression fits using the following relation:
\begin{equation}
    \log(\frac{\tau_{\rm rest}}{{\rm s}}) = A \log\left(\frac{M_{\rm BH}}{M_\odot}\right) 
    + B\Gamma 
    + C \log\left(\frac{L_{\rm X}}{\rm erg\ s^{-1}}\right)+ D+\epsilon,
\end{equation}
where the coefficients $A$, $B$, $C$, and $D$ represent the dependencies and the intercept term, respectively. 
{And $\epsilon \sim \mathcal{N}(0, \sigma_\epsilon^2)$ accounts for the uncertainties in the relation.}
We considered all possible parameter combinations among black hole mass $M_{\rm BH}$, 
X-ray photon index $\Gamma$, and X-ray 2--10~keV luminosity $L_{\rm X}$. 
For each configuration, we ran the MCMC fitting with 10,000 sampling steps and computed the Bayesian Information Criterion (BIC) as ${\rm BIC} = -2\ln \mathcal{L} + N\ln N_{\rm data}$, where $\mathcal{L}$ is the likelihood, $N$ is the number of free parameters, and $N_{\rm data}$ is the number of data points.
The likelihood was evaluated at the median values of the probability distribution for each model.
The BIC values were then compared to assess the relative performance of different models, 
with smaller BIC values indicating a better balance between goodness of fit and model complexity.
For each fitted parameter, we report the median value along with the 1$\sigma$ uncertainties derived from the MCMC.
The resulting coefficients and BIC values for all seven models are summarized in Table~\ref{tab1}.

\begin{figure}[t]
    \centering
    \includegraphics[width=1\linewidth]{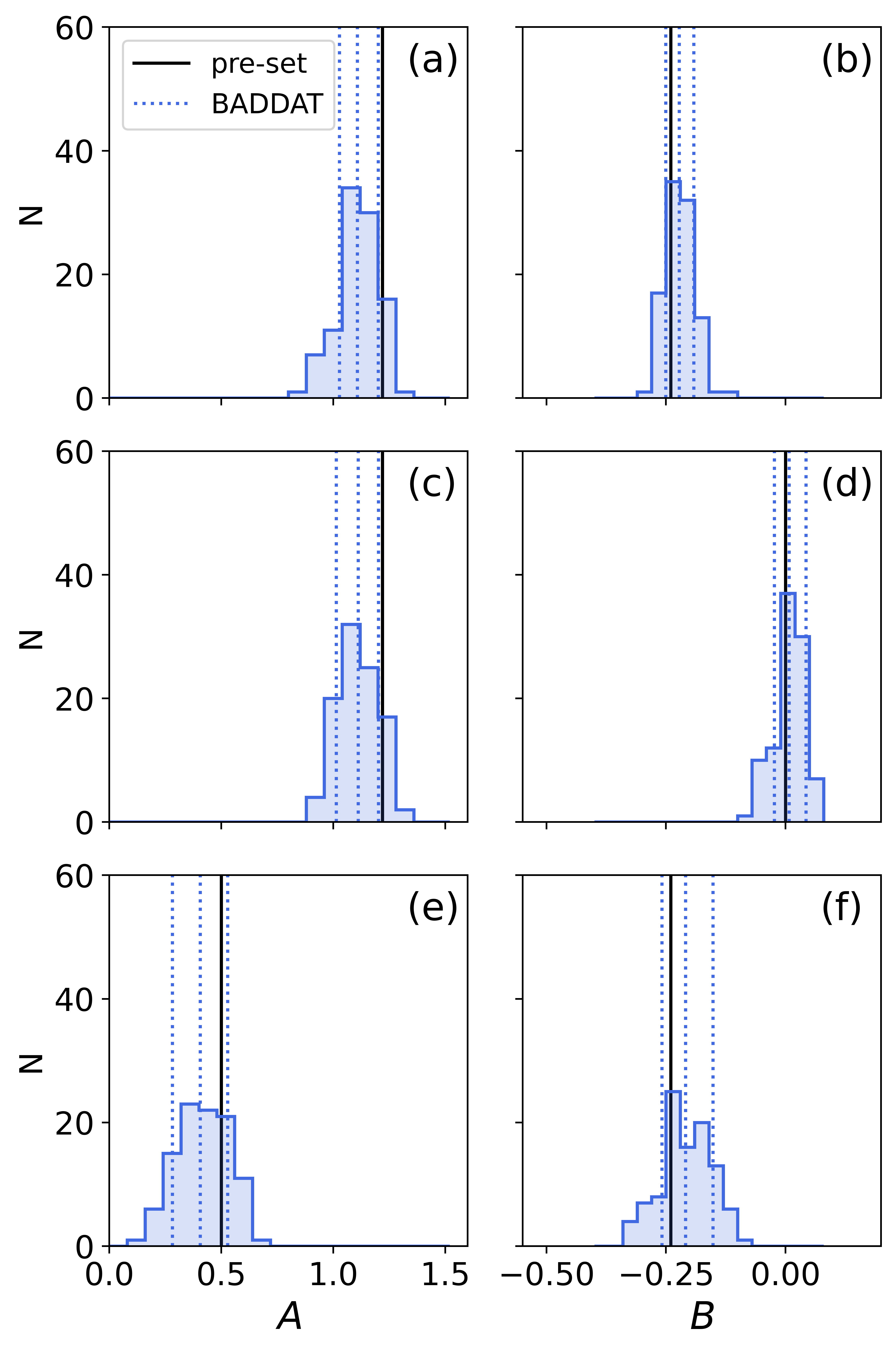}
    \caption{The distributions of the fitted coefficients $A$ and $B$ for the timescale dependence on $\log(M_{\rm BH}/M_\odot)$ and $\Gamma$ on the mock sample. 
    {The vertical black solid line indicates the preset coefficient, while the blue dotted lines show the 16th, 50th, and 84th percentiles of the distribution obtained from BADDAT.}
    }
    \label{fig3}
\end{figure}

It is evident that the model including both $M_{\rm BH}$ and $\Gamma$ (Model 4) achieves the lowest BIC value, suggesting that it is statistically the most favored model.
The corresponding coefficients are \(A = 1.22^{+0.11}_{-0.09}\) and \(B = -0.24^{+0.06}_{-0.07}\), indicating a positive correlation of the variability timescale with black hole mass and a weak negative (but significant) dependence on the photon index.  
Models including $ L_{\rm X}$ do not significantly improve the BIC. 
{This is likely because its effect is largely subsumed by $M_{\rm BH}$, given the strong correlation between $L_{\rm X}$ and $M_{\rm BH}$, with $M_{\rm BH}$ playing the dominant role and $L_{\rm X}$ contributing only secondarily, as also discussed in Section~\ref{discussion_timescale_dependence}.}
Therefore, we adopt this relation (Model 4) as our estimated variability plane, with the sample and the corresponding fit illustrated in Figure~\ref{fig2}.
{Data points with modeled intrinsic timescale in the observer frame $\tau_{\rm in}=A\log(M_{\rm BH}/M_\odot)+B\Gamma+D+\log(1+z)$ shorter than 0.1$\times$baseline (i.e., $\rho_{\rm in}<0.1$) are shown as dots and the others (i.e., $\rho_{\rm in}\geq0.1$) as triangles for visual reference, as they are likely to be underestimated.}
BADDAT does not rely on this threshold, but fully accounts for the probability distributions of all data points, yielding reliable regression results \citep{2025MNRAS.544L..96X}, as will be specifically demonstrated by recovering hypothesis correlations on mock light curves for our X-ray sample in Section~\ref{sec_test}.
{
Because we adopt uniform priors on the parameters (in log space), the resulting distributions from EMCEE in the right panel of Figure~\ref{fig2} reflect the likelihood distributions.
Note that there is a clear degeneracy between the intercept parameter, $D$, and the coefficient, $A$.
A mean subtraction could mitigate the degeneracy, but we confirm that the results, including the values and uncertainties of the coefficients and the intercept, remain largely unchanged if it is done.
}

\section{Discussions}\label{sec_dis}
\subsection{Recovering Hypothesis Correlations on Mock Light Curves} \label{sec_test}

As recommended by \citet{2025MNRAS.544L..96X}, the effectiveness of BADDAT strongly depends on the quality of the samples. 
Therefore, a simple mock test is required to verify that BADDAT is appropriate and nearly unbiased for this X-ray sample.
{The mock light curves are generated following the procedure in \citet{2025MNRAS.544L..96X}, using \textsc{celerite} as described in \citet{burke_characteristic_2021}.
Note that we have confirmed their consistency in characterizing DRW variability by generating light curves using both \textsc{celerite} in Python and \texttt{arima.sim} in \textsc{R} \citep{R_2023} with the same set of parameters.
}

For each of the three assumed input relations described below, we generated 100 realizations of mock samples.
The first relation adopts the best-fit result from our regression in Section~\ref{sec_BADDAT}:
\begin{equation}
\log\tau_{\rm rest} = 1.22\log \frac{M_{\rm BH}}{M_\odot} - 0.24\Gamma - 3.53.
\end{equation}
In the second case, we fix the coefficient of $\Gamma$ to zero and account for its effect using the mean $\Gamma$ value of the sample:
\begin{equation}
\log\tau_{\rm rest} = 1.22\log \frac{M_{\rm BH}}{M_\odot} - 3.99.
\end{equation}
In the third case, we reduce the coefficient of $\log M_{\rm BH}$ to 0.5 and account for the remaining $0.72\log M_{\rm BH}$ contribution using its mean value:
\begin{equation}
\log\tau_{\rm rest} = 0.5\log \frac{M_{\rm BH}}{M_\odot} - 0.24\Gamma + 1.59.
\end{equation}

Figure~\ref{fig3} demonstrates the performance of BADDAT on this X-ray sample.
For each of the three assumed input relations, we compare the recovered coefficients $A$ and $B$ in 
$\log(\tau_{\rm rest}/{\rm s}) = A\log\frac{M_{\rm BH}}{M_\odot} + B\Gamma + D$ with their pre-set values. 
The recovered coefficients $A$ for $M_{\rm BH}$ are slightly biased toward lower values relative to the preset inputs, likely due to the limited number of independent sources, as multiple exposures correspond to the same sources and the $M_{\rm BH}$ estimates are therefore not fully independent (i.e., only 112 $M_{\rm BH}$ measurements in a sample containing 399 light curves).
Nevertheless, the distributions correctly track the input values when they are varied, demonstrating that BADDAT is sensitive to the underlying parameter dependencies, with deviations generally $\lesssim 1\sigma$ of the distributions.
A larger sample with more black hole mass measurements is required to achieve better constraints on the $\tau$–$M_{\rm BH}$ relation.
In contrast, each exposure provides an independent estimate of $\Gamma$, which accounts for the accuracy of the recovered coefficients $B$.

\subsection{Timescale Dependences}\label{discussion_timescale_dependence}

%A
Our resulting X-ray timescale dependence on black hole mass ($A=1.22$) is consistent with previous measurements \citep[e.g.][]{gonzalezmartin_xray_2012, 2025arXiv251014529L}, despite using a different method to estimate the timescale.
As shown by the BIC values in Table~\ref{tab1}, the models considering $M_{\rm BH}$ ({all having} $\text{BIC}<-200$) are significantly better fits to the sample than those without $M_{\rm BH}$ ({having} $\text{BIC}\gtrsim -100$), indicating that it is $M_{\rm BH}$ that dominates the timescale {dependence} of AGN X-ray variability.

% B
As the regression results suggest, a softer spectrum is associated with faster X-ray variability.
This finding is not due to the variance of dominant energy bands, since a higher photon index $\Gamma$ corresponds to a softer spectrum, where the emission is dominated by soft X-rays that are generally expected to vary more slowly than the hard X-rays.
Moreover, this trend cannot be attributed to the known correlation between $\Gamma$ and $L_{\rm X}$: a larger $\Gamma$ typically indicates a higher $L_{\rm X}$, which in turn would suggest slower variability.
Therefore, this dependence on the photon index $\Gamma$ can be intrinsic, indicating that the variability timescale correlates with the structure of the corona.
Our univariate result ({Model} 2 in Table~\ref{tab1}), $\log (\tau_{\rm rest}/{\rm s} )= -0.76^{+0.10}_{-0.11}\Gamma + 5.88^{+0.24}_{-0.21}$, is consistent with the result $\tau=-(43.3\pm6.2){\rm ksec}\cdot\Gamma + (85.9\pm10.6){\rm ksec}$ from \citet{1997A&A...326L..25K} when linearly approximated about $\Gamma=2$.
They proposed that Comptonization could account for the observed relationship, as a larger number of Compton collisions is required to produce the high-energy photons in harder spectra (smaller photon {indexes}), thereby leading to longer variability timescales. 
This scenario can be supported by the expected inverse correlation between the photon index and the Compton optical depth \citep{1994ApJ...434..570T}.

% C
{Luminosity} has previously been considered as a factor in the variability plane. {However,} it is correlated with black hole mass. 
BIC analysis indicates that once black hole mass is included, adding luminosity does not improve the model. 
This suggests that the apparent dependence of $\tau_{\rm rest}$ on $L_{\rm X}$ primarily reflects the underlying correlation between $L_{\rm X}$ and $M_{\rm BH}$ and may not provide additional independent information. 
This result is consistent with \citet{gonzalezmartin_xray_2012} and \citet{2018ApJ...858....2G}, who also reported that luminosity is not required in the X-ray variability plane.
Additional regressions considering the {Eddington ratio} ($\dot{m}$) are presented in Appendix~\ref{appendixB}, along with further discussions.

\subsection{Is DRW Effective for Characterizing X-ray Variability Timescale?}

A universal shape for the power spectrum of quasars in the optical/UV bands, described by a broken power-law, has been established \citep[e.g.,][]{2024A&A...684A.133A, 2024A&A...686A.286P, 2025A&A...701A.295P} and modeled \citep[e.g.,][]{2018ApJ...855..117C, 2020ApJ...892...63C, 2020ApJ...902....7S}.
Although exceptions exist \citep[e.g.,][]{2011ApJ...743L..12M, 2024ApJ...969...78S}, optical light curves of AGNs can be roughly described by the DRW model, whose power spectral density corresponds to a broken power-law with slopes of $0$ and $-2$ at low and high frequencies, respectively.
Strong correlations have been identified between the optical damping timescale and AGN properties such as black hole mass, bolometric luminosity, and wavelength \citep[e.g.,][]{burke_characteristic_2021, 2024ApJ...966....8Z, 2024ApJ...975..160R, 2025MNRAS.544L..96X}.
In contrast, X-ray variability exhibits a more complex behavior, cautioning that the DRW model may not always provide a robust description in this regime \citep[e.g.,][]{2019MNRAS.482.2088A, 2019MNRAS.485..260A, 2025MNRAS.539.1775L}.
Nevertheless, the X-ray characteristic break timescale has proven to be an effective tracer of correlations with AGN properties \citep[e.g.,][]{markowitz_xray_2003, mchardy_Active_2006, gonzalezmartin_xray_2012, yang_exploring_2022}, motivating the use of the DRW process, which characterizes variability with a comparable break timescale.

In this work, we adopt the DRW model as a simplified approach to characterize variability using a single parameter, the damping timescale $\tau$.
{This may introduce biases in the inferred timescales. 
However, our analysis focuses on population-level regression rather than on the precise characterization of the PSD for individual sources. 
In this context, the DRW timescale can still serve as a useful statistical quantity that captures the overall variability properties of AGNs.
A population-level correlation is identified by BADDAT and robustly validated through mock simulations. 
Any biases introduced by fitting a simple DRW model are expected to contribute to the uncertainties in the BADDAT regression rather than to affect the overall relation.}

{As discussed in \citet{2019MNRAS.485..260A}, the PSD can be considered approximately stationary on $\lesssim$ days timescales. 
The individual exposures in our sample are basically continuous, with a maximum duration of about one day. 
Therefore, non-stationarity does not affect the variability modeling within a single exposure. 
However, such non-stationarity across different exposures of a given source could naturally contribute to the uncertainties in the BADDAT regression.
}

Therefore, we conclude that the DRW model provides an effective description of X-ray variability at the level of precision required in this study, where a single characteristic timescale, $\tau$, is sufficient to capture the variability responsible for the observed population-level correlation.
However, a more sophisticated timescale fitting using a more appropriate Gaussian process kernel \citep[e.g.,][]{2022MNRAS.514..164S, 2025MNRAS.539.1775L, 2025ApJ...984...45X} would benefit further detailed analyses of the X-ray variability plane.

\section{Summary}\label{sec_con}
In this work, we investigated the X-ray variability properties of AGNs and established an updated variability plane that incorporates the photon index. The conclusions are detailed as follows:

\begin{enumerate}
\item We compile a sample of 112 AGNs with 399 exposures from the 4XMM-DR14 {catalog} with {type and redshift information} from SIMBAD and black hole mass measurements from the literature.

\item We find {that} the X-ray {rest-frame} variability timescale $\tau_{\rm rest}$ of these AGNs is correlated with the black hole mass $M_{\rm BH}$, photon index $\Gamma$ and X-ray luminosity $L_{\rm X}$. However, through BIC analysis, we find that the apparent dependence of $\tau_{\rm rest}$ on $L_{\rm X}$ primarily reflects the underlying correlation between $L_{\rm X}$ and $M_{\rm BH}$. This suggests that $M_{\rm BH}$ is the dominant parameter governing the variability timescale.

\item Applying the recently developed fitting method BADDAT, we confirm the dependence of $\tau_{\rm rest}$ on $M_{\rm BH}$ and further incorporate $\Gamma$ into the variability plane, yielding a best-fit relation of 
$\log (\tau_{\rm rest}/{\rm s})=1.22^{+0.11}_{-0.09}\log (M_{\rm BH}/M_\odot) - 0.24^{+0.06}_{-0.07}\Gamma - 3.53^{+0.62}_{-0.73}$, which is strongly favored by the BIC.

\end{enumerate}

\begin{acknowledgments}
The authors gratefully acknowledge Zhen-Yi Cai for the original idea of the BADDAT method in our previous work, which made this work possible.
This research has made use of data obtained from the 4XMM XMM-Newton Serendipitous Source Catalog compiled by the 10 institutes of the XMM-Newton Survey Science Centre selected by ESA.
This research has made use of the SIMBAD database, operated at CDS, Strasbourg, France.
{R.S.X., H.L., Y.Q.X, J.L.W, G.W.R, S.F.Z \& M.Q.H acknowledge support from the National Key R\&D Program of China (2023YFA1608100 and 2022YFF0503401), the Strategic Priority Research Program of the Chinese Academy of Sciences (grant NO. XDB0550300), and the NSFC grants (12025303 and 12393814).}
G.W.R acknowledges support from the Anhui Provincial Natural Science Foundation  (2508085QA007) and the China Postdoctoral Science Foundation (grant No. 2025M773191).

\end{acknowledgments}

\bibliography{sample631}{}

\appendix

\renewcommand{\thetable}{A\arabic{table}}
\setcounter{table}{0}

\section{Detailed Information on Final {Sample}}\label{appendixA}

Table~\ref{tabA} presents detailed information on the sources included in our final sample.

{
\renewcommand{\small}{\fontsize{7}{7}\selectfont}
\vspace{0.1cm}
\setlength{\LTcapwidth}{17cm}
\begin{longtable*}{|l|l|r|r|r|c|l|l|l|}
\caption{Information about our final sample. 
{We list the source names, types, coordinates, and redshifts from SIMBAD, along with the black hole mass estimates, the estimation methods, the corresponding references, {and the 4XMM-DR14 exposures}.
}
The methods are noted:
SE: single epoch;
$\sigma$: $M$-$\sigma$ relation;
DM: dynamic modeling;
RM: reverberation mapping;
B: $M$-Bulge mass or luminosity relation;
QS: QPO mass scaling.
{$^*$The extended list of exposures for all sources in our sample is provided in the machine-readable version of this table, available online.}
}
\phantomsection
\label{tabA}
\\

\hline
Name & Type & RA & DEC & Redshift  & Method & $\log M_{\rm BH}$& Reference &  4XMM-DR14 \\
&&(deg.)&(deg.)&&&($M_\odot$)&&Exposure(s)$^*$ \\
\hline
\endfirsthead

\hline
Name & Type & RA & DEC & Redshift  & Method & $\log M_{\rm BH}$& Reference &  4XMM-DR14 \\
&&(deg.)&(deg.)&&&($M_\odot$)&&Exposure(s)$^*$ \\
\hline
\endhead

\hline
\endfoot
\hline
\endlastfoot
UGC 12163 & Sy1 &  340.6639 & 29.7253 &0.0243& SE & $6.53 \pm 0.10$ & \text{\citet{2005ApJ...625...78H}} & EPN\_S003, etc. \\
LBQS 1244+0238 & Sy1 &  191.6468 & 2.3690 &0.0486& SE & $6.30 \pm 0.10$ & \text{\citet{2007ApJ...658..815S}} & EPN\_S003, etc. \\
Mrk 279 & Sy1 &  208.2642 & 69.3085 &0.0302& SE & $7.60 \pm 0.10$ & \text{\citet{2007ApJ...658..815S}} & EPN\_S003 \\
LEDA 17155 & Sy2 &  80.2558 & -25.3626 &0.0409& $\sigma$ & $7.86 \pm 0.10$ & \text{\citet{2010ApJ...720..786L}} & EPN\_S003 \\
IRAS 13349+2438 & Sy1 &  204.3280 & 24.3843 &0.1083& $\sigma$ & $9.00 \pm 0.10$ & \text{\citet{2013MNRAS.430.2650L}} & EPN\_S003, etc. \\
NGC 4253 & Sy1 &  184.6104 & 29.8130 &0.0129& SE & $6.60 \pm 0.10$ & \text{\citet{2007ApJ...658..815S}} & EPN\_S011, etc. \\
Mrk 335 & Sy1 &  1.5814 & 20.2030 &0.0259& SE & $7.23 \pm 0.10$ & \text{\citet{2018A&A...614A.120S}} & EPN\_S001, etc. \\
ESO 113$-$45 & Sy1 &  20.9405 & -58.8057 &0.0459& SE & $7.90 \pm 0.10$ & \text{\citet{2007ApJ...658..815S}} & EPN\_S003 \\
NGC 3227 & Sy1 &  155.8773 & 19.8651 &0.0033& RM & $7.35 \pm 0.23$ & \text{\citet{2013ApJ...773...90G}} & EPN\_U002, etc. \\
ESO 445$-$50 & Sy1 &  207.3304 & -30.3094 &0.0160& RM & $7.83 \pm 0.07$ & \text{\citet{2023ApJ...944...29B}} & EPN\_S003, etc. \\
ESO 141$-$55 & Sy1 &  290.3087 & -58.6702 &0.0371& $\sigma$ & $7.60 \pm 0.24$ & \text{\citet{2016MNRAS.458.2454L}} & EPN\_S003, etc. \\
2MASS J08105865+7602424 & Sy1 &  122.7444 & 76.0452 &0.0988& RM & $8.24 \pm 0.10$ & \text{\citet{2002ApJ...579..530W}} & EPN\_S003 \\
Mrk 1383 & Sy1 &  217.2774 & 1.2850 &0.0859& SE & $8.72 \pm 0.10$ & \text{\citet{2005ApJ...625...78H}} & EPN\_S003 \\
LEDA 87814 & Sy1 &  14.6557 & -36.1013 &0.1641& SE & $8.72 \pm 0.10$ & \text{\citet{2003A&A...408..119P}} & EPN\_S003 \\
Ton 951 & Sy1 &  131.9269 & 34.7512 &0.0640& SE & $7.86 \pm 0.10$ & \text{\citet{2018A&A...614A.120S}} & EPN\_S001 \\
Mrk 79 & Sy1 &  115.6371 & 49.8100 &0.0221& SE & $7.61 \pm 0.10$ & \text{\citet{2018A&A...614A.120S}} & EPN\_S003, etc. \\
LEDA 3096673 & Sy1 &  334.8272 & 12.1315 &0.0813& SE & $6.30 \pm 0.10$ & \text{\citet{2016MNRAS.457..875P}} & EPN\_S001 \\
ESO 113$-$10 & Sy1 &  16.3202 & -58.4375 &0.0260& SE & $6.85 \pm 0.10$ & \text{\citet{2013ApJ...764L...9C}} & EPN\_S003 \\
NGC 3516 & Sy1 &  166.6978 & 72.5688 &0.0087& SE & $7.39 \pm 0.10$ & \text{\citet{2018A&A...614A.120S}} & EPN\_S027, etc. \\
Mrk 478 & Sy1 &  220.5312 & 35.4397 &0.0777& SE & $7.23 \pm 0.10$ & \text{\citet{2005ApJ...625...78H}} & EPN\_S003, etc. \\
Z 212$-$25 & Sy1 &  158.6607 & 39.6413 &0.0431& QS & $6.48 \pm 0.10$ & \text{\citet{2018MNRAS.478.4830C}} & EPN\_S003, etc. \\
2MASS J22484115$-$5109532 & Sy1 &  342.1714 & -51.1647 &0.1024& SE & $8.10 \pm 0.10$ & \text{\citet{2014MNRAS.437.3929S}} & EPN\_S001 \\
2MASS J13234951+6541480 & Sy1 &  200.9570 & 65.6965 &0.1678& SE & $8.29 \pm 0.10$ & \text{\citet{2012ApJS..201...38T}} & EPN\_S003 \\
PB 4142 & Sy1 &  208.6487 & 18.0882 &0.1515& SE & $8.35 \pm 0.06$ & \text{\citet{2022ApJS..263...42W}} & EPN\_S003 \\
Ton 182 & Sy1 &  211.3176 & 25.9261 &0.1640& SE & $7.70 \pm 0.10$ & \text{\citet{2005ApJ...625...78H}} & EPN\_S003 \\
NGC 526 & Sy2 &  20.9763 & -35.0654 &0.0189& RM & $7.60 \pm 0.10$ & \text{\citet{2012ApJ...745..107W}} & EPN\_S003, etc. \\
Mrk 1506 & Sy1 &  68.2962 & 5.3544 &0.0331& RM & $7.72 \pm 0.04$ & \text{\citet{2013ApJ...773...90G}} & EPN\_S003, etc. \\
NGC 4051 & Sy1 &  180.7900 & 44.5313 &0.0020& RM & $6.34 \pm 0.08$ & \text{\citet{2013ApJ...773...90G}} & EPN\_S003, etc. \\
NGC 5548 & Sy1 &  214.4981 & 25.1369 &0.0167& SE & $7.72 \pm 0.10$ & \text{\citet{2018A&A...614A.120S}} & EPN\_S003, etc. \\
NGC 4593 & Sy1 &  189.9146 & -5.3443 &0.0083& SE & $6.88 \pm 0.10$ & \text{\citet{2018A&A...614A.120S}} & EPN\_S003, etc. \\
LEDA 88835 & Sy1 &  201.3306 & -38.4149 &0.0658& SM & $6.54 \pm 0.10$ & \text{\citet{2015MNRAS.446..759C}} & EPN\_S002, etc. \\
LEDA 88588 & Sy1 &  107.1727 & -49.5518 &0.0406& SE & $6.31 \pm 0.10$ & \text{\citet{2003MNRAS.343..164B}} & EPN\_S003, etc. \\
Mrk 1502 & Sy1 &  13.3955 & 12.6933 &0.0612& RM & $6.96 \pm 0.06$ & \text{\citet{2019ApJ...876..102H}} & EPN\_U002, etc. \\
Ton S 180 & Sy1 &  14.3343 & -22.3826 &0.0617& SE & $7.09 \pm 0.10$ & \text{\citet{2005ApJ...625...78H}} & EPN\_S004, etc. \\
HE 1029$-$1401 & Sy1 &  157.9761 & -14.2807 &0.0852& $\sigma$ & $8.70 \pm 0.30$ & \text{\citet{2010A&A...519A.115H}} & EPN\_S001 \\
PG 0953+414 & Sy1 &  149.2183 & 41.2562 &0.2341& RM & $8.44 \pm 0.10$ & \text{\citet{2004ApJ...613..682P}} & EPN\_S003 \\
Ton 1388 & Sy1 &  169.7861 & 21.3216 &0.1760& SE & $8.49 \pm 0.09$ & \text{\citet{2022ApJS..263...42W}} & EPN\_S003, etc. \\
ESO 383$-$35 & Sy1 &  203.9741 & -34.2955 &0.0071& $\sigma$ & $6.68 \pm 0.10$ & \text{\citet{2005MNRAS.359.1469M}} & EPN\_S003, etc. \\
NGC 7314 & Sy2 &  338.9426 & -26.0504 &0.0046& SE & $6.24 \pm 0.06$ & \text{\citet{2017MNRAS.468L..97O}} & EPN\_S003, etc. \\
NGC 7213 & Sy1 &  332.3176 & -47.1665 &0.0048& $\sigma$ & $7.90 \pm 0.50$ & \text{\citet{2014MNRAS.438.3322S}} & EPN\_S009 \\
NGC 3783 & Sy1 &  174.7573 & -37.7388 &0.0090& RM & $7.28 \pm 0.09$ & \text{\citet{2013ApJ...773...90G}} & EPN\_S003, etc. \\
NGC 4151 & Sy1 &  182.6357 & 39.4059 &0.0032& SE & $7.56 \pm 0.10$ & \text{\citet{2018A&A...614A.120S}} & EPN\_S001, etc. \\
NGC 4395 & Sy2 &  186.4536 & 33.5467 &0.0011& RM & $5.56 \pm 0.10$ & \text{\citet{2005ApJ...632..799P}} & EPN\_S003, etc. \\
NGC 5273 & Sy1 &  205.5349 & 35.6543 &0.0036& $\sigma$ & $6.51 \pm 0.10$ & \text{\citet{2002ApJ...579..530W}} & EPN\_S003, etc. \\
Mrk 1044 & Sy1 &  37.5230 & -8.9981 &0.0173& SE & $6.34 \pm 0.10$ & \text{\citet{2005ApJ...625...78H}} & EPN\_S003, etc. \\
LB 1727 & Sy1 &  66.5029 & -57.2004 &0.1041& RM & $8.70 \pm 0.10$ & \text{\citet{2012ApJ...745..107W}} & EPN\_S003, etc. \\
Mrk 359 & Sy1 &  21.8855 & 19.1786 &0.0168& SE & $6.46 \pm 0.10$ & \text{\citet{2014MNRAS.438.2828D}} & EPN\_S003, etc. \\
Mrk 493 & Sy1 &  239.7901 & 35.0299 &0.0310& SE & $6.30 \pm 0.10$ & \text{\citet{2010ApJS..187...64G}} & EPN\_S003, etc. \\
PB 3894 & Sy1 &  183.5737 & 14.0537 &0.0809& RM & $7.49 \pm 0.10$ & \text{\citet{2002ApJ...579..530W}} & EPN\_S003, etc. \\
ESO 434$-$40 & Sy2 &  146.9176 & -30.9488 &0.0084& SE & $7.22 \pm 0.10$ & \text{\citet{2017MNRAS.468L..97O}} & EPN\_S003, etc. \\
ESO 198$-$24 & Sy1 &  39.5821 & -52.1923 &0.0453& RM & $8.10 \pm 0.10$ & \text{\citet{2012ApJ...745..107W}} & EPN\_S003, etc. \\
Mrk 841 & Sy1 &  226.0050 & 10.4377 &0.0366& SE & $8.10 \pm 0.10$ & \text{\citet{2007ApJ...658..815S}} & EPN\_S003, etc. \\
NAME MR 2251$-$178 & Sy1 &  343.5245 & -17.5820 &0.0645& RM & $8.50 \pm 0.10$ & \text{\citet{2012ApJ...745..107W}} & EPN\_S003, etc. \\
2MASS J05594739$-$5026519 & Sy1 &  89.9472 & -50.4477 &0.1375& SE & $7.76 \pm 0.10$ & \text{\citet{2010A&A...510A..65P}} & EPN\_S001, etc. \\
Mrk 509 & Sy1 &  311.0407 & -10.7234 &0.0347& RM & $7.98 \pm 0.02$ & \text{\citet{2013ApJ...773...90G}} & EPN\_S003, etc. \\
Mrk 1095 & Sy1 &  79.0475 & -0.1498 &0.0326& SE & $8.07 \pm 0.10$ & \text{\citet{2018A&A...614A.120S}} & EPN\_S003, etc. \\
NGC 7172 & Sy2 &  330.5078 & -31.8696 &0.0085& $\sigma$ & $7.67 \pm 0.10$ & \text{\citet{2010ApJ...720..786L}} & EPN\_S001 \\
2MASSI J0918486+211717 & Sy1 &  139.7025 & 21.2880 &0.1490& SE & $7.37 \pm 0.02$ & \text{\citet{2022ApJS..263...42W}} & EPN\_S003 \\
NGC 985 & Sy1 &  38.6577 & -8.7878 &0.0430& B & $8.94 \pm 0.10$ & \text{\citet{2009ApJ...690.1322W}} & EPN\_S003, etc. \\
Mrk 1513 & Sy1 &  323.1160 & 10.1386 &0.0610& DM & $6.92 \pm 0.24$ & \text{\citet{2017ApJ...849..146G}} & EPN\_S003 \\
2MASX J14510879+2709272 & Sy1 &  222.7865 & 27.1575 &0.0645& SE & $7.46 \pm 0.04$ & \text{\citet{2022ApJS..263...42W}} & EPN\_S003, etc. \\
ATO J176.4186$-$18.4541 & Sy1 &  176.4187 & -18.4542 &0.0326& SE & $7.60 \pm 0.10$ & \text{\citet{2020A&A...634A..92U}} & EPN\_S001, etc. \\
Mrk 110 & Sy1 &  141.3033 & 52.2861 &0.0352& SE & $7.29 \pm 0.10$ & \text{\citet{2018A&A...614A.120S}} & EPN\_S003 \\
UGC 3374 & Sy1 &  88.7238 & 46.4400 &0.0202& SE & $8.07 \pm 0.10$ & \text{\citet{2010ApJ...710..503W}} & EPN\_S009 \\
Mrk 1298 & Sy1 &  172.3195 & -4.4022 &0.0600& $\sigma$ & $8.08 \pm 0.10$ & \text{\citet{2007ApJ...657..102D}} & EPN\_S003 \\
LEDA 42648 & Sy1 &  190.5442 & 33.2841 &0.0435& RM & $6.80 \pm 0.27$ & \text{\citet{2014ApJ...782...45D}} & EPN\_U002, etc. \\
LEDA 3096712 & Sy1 &  344.4126 & -36.9351 &0.0390& SE & $6.59 \pm 0.10$ & \text{\citet{2010ApJS..187...64G}} & EPN\_S003, etc. \\
6C 170204+454510 & Sy1 &  255.8766 & 45.6798 &0.0607& SE & $6.77 \pm 0.10$ & \text{\citet{2001A&A...377...52W}} & EPN\_S003, etc. \\
Mrk 704 & Sy1 &  139.6083 & 16.3055 &0.0295& SE & $8.11 \pm 0.10$ & \text{\citet{2009AJ....137.3388W}} & EPN\_S003 \\
LEDA 26550 & Sy1 &  140.6960 & 51.3439 &0.1597& SE & $8.02 \pm 0.27$ & \text{\citet{2022ApJS..263...42W}} & EPN\_S003 \\
MCG$-$03$-$58$-$007 & Sy2 &  342.4048 & -19.2740 &0.0319& $\sigma$ & $8.00 \pm 0.10$ & \text{\citet{2018MNRAS.479.3592B}} & EPN\_S003, etc. \\
RX J0136.9$-$3510 & Sy1 &  24.2268 & -35.1645 &0.2890& SE & $7.90 \pm 0.10$ & \text{\citet{2009MNRAS.398L..16J}} & EPN\_S003 \\
2MASX J11400874+0307114 & Sy1 &  175.0363 & 3.1198 &0.0810& SE & $5.77 \pm 0.10$ & \text{\citet{2004ApJ...610..722G}} & EPN\_S003, etc. \\
ESO 548$-$81 & Sy1 &  55.5155 & -21.2443 &0.0144& RM & $8.60 \pm 0.10$ & \text{\citet{2012ApJ...745..107W}} & EPN\_S001 \\
ESO 362$-$18 & Sy2 &  79.8990 & -32.6577 &0.0125& RM & $8.70 \pm 0.10$ & \text{\citet{2012ApJ...745..107W}} & EPN\_S001, etc. \\
NGC 6860 & Sy2 &  302.1955 & -61.0998 &0.0148& RM & $7.80 \pm 0.10$ & \text{\citet{2012ApJ...745..107W}} & EPN\_S002 \\
Mrk 290 & Sy1 &  233.9682 & 57.9026 &0.0304& SE & $7.90 \pm 0.10$ & \text{\citet{2010ApJ...710..503W}} & EPN\_S003 \\
NGC 6221 & Sy2 &  253.1929 & -59.2168 &0.0041& SE & $6.46 \pm 0.10$ & \text{\citet{2017MNRAS.468L..97O}} & EPN\_U005, etc. \\
QSO B1725$-$142 & QSO &  262.0825 & -14.2653 &0.1840& SE & $8.66 \pm 0.10$ & \text{\citet{2009AJ....137.3388W}} & EPN\_S003, etc. \\
ESO 511$-$30 & Sy1 &  214.8432 & -26.6448 &0.0229& B & $8.66 \pm 0.10$ & \text{\citet{2009ApJ...690.1322W}} & EPN\_U002 \\
2MASS J01341690$-$4258262 & Sy1 &  23.5703 & -42.9739 &0.2371& SE & $7.17 \pm 0.10$ & \text{\citet{2010ApJS..187...64G}} & EPN\_S003, etc. \\
LEDA 90334 & Sy1 &  294.3876 & -6.2180 &0.0104& SE & $6.48 \pm 0.10$ & \text{\citet{2008MNRAS.389.1360M}} & EPN\_S003, etc. \\
NGC 6814 & Sy1 &  295.6690 & -10.3236 &0.0058& RM & $6.42 \pm 0.24$ & \text{\citet{2014MNRAS.445.3073P}} & EPN\_S001, etc. \\
NGC 931 & Sy1 &  37.0602 & 31.3114 &0.0163& $\sigma$ & $7.64 \pm 0.10$ & \text{\citet{2002ApJ...579..530W}} & EPN\_S001, etc. \\
NGC 3660 & Sy2 &  170.8844 & -8.6585 &0.0123& SE & $7.08 \pm 0.24$ & \text{\citet{2012MNRAS.426.3225B}} & EPN\_S003 \\
IRAS 21262+5643 & Sy1 &  321.9395 & 56.9430 &0.0149& SE & $7.18 \pm 0.10$ & \text{\citet{2008MNRAS.389.1360M}} & EPN\_S003, etc. \\
Mrk 817 & Sy1 &  219.0920 & 58.7943 &0.0312& SE & $7.59 \pm 0.10$ & \text{\citet{2018A&A...614A.120S}} & EPN\_S003, etc. \\
2MASS J07511218+1743517 & Sy1 &  117.8008 & 17.7310 &0.1861& SE & $7.90 \pm 0.18$ & \text{\citet{2022ApJS..263...42W}} & EPN\_S003 \\
Mrk 382 & Sy1 &  118.8555 & 39.1862 &0.0332& SE & $6.71 \pm 0.10$ & \text{\citet{2005ApJ...625...78H}} & EPN\_S001, etc. \\
2MASX J19271951+6533539 & Sy1 &  291.8317 & 65.5653 &0.0170& B & $5.98 \pm 0.12$ & \text{\citet{2022ApJ...933...70L}} & EPN\_S002, etc. \\
Z 229$-$15 & Sy1 &  286.3581 & 42.4611 &0.0276& RM & $7.00 \pm 0.12$ & \text{\citet{2011ApJ...732..121B}} & EPN\_S003 \\
Mrk 1310 & Sy1 &  180.3098 & -3.6781 &0.0195& RM & $7.42 \pm 0.10$ & \text{\citet{2014MNRAS.445.3073P}} & EPN\_S003 \\
NGC 2617 & Sy1 &  128.9116 & -4.0883 &0.0143& RM & $7.45 \pm 0.10$ & \text{\citet{2018ApJ...854..107F}} & EPN\_S001 \\
NGC 4748 & Sy1 &  193.0522 & -13.4148 &0.0141& RM & $6.48 \pm 0.16$ & \text{\citet{2013ApJ...773...90G}} & EPN\_S007 \\
LEDA 801745 & Sy2 &  174.7129 & -23.3598 &0.0271& RM & $7.58 \pm 0.10$ & \text{\citet{2018A&A...619A.168K}} & EPN\_S003 \\
QSO J0439$-$5311 & Sy1 &  69.9110 & -53.1920 &0.2430& SE & $6.59 \pm 0.10$ & \text{\citet{2010ApJS..187...64G}} & EPN\_S003, etc. \\
LEDA 2816068 & Sy1 &  208.8189 & 56.2125 &0.1215& SE & $7.35 \pm 0.22$ & \text{\citet{2022ApJS..263...42W}} & EPN\_S003, etc. \\
Mrk 915 & Sy2 &  339.1938 & -12.5452 &0.0239& SE & $7.76 \pm 0.37$ & \text{\citet{2021MNRAS.506.4960H}} & EPN\_S003, etc. \\
LEDA 89420 & Sy2 &  351.3507 & -38.4474 &0.0361& SE & $8.23 \pm 0.10$ & \text{\citet{2010ApJS..187...64G}} & EPN\_S003 \\
2MASS J08010140+1848409 & Sy1 &  120.2558 & 18.8113 &0.1395& SE & $7.57 \pm 0.30$ & \text{\citet{2022ApJS..263...42W}} & EPN\_S003 \\
NGC 1566 & Sy1 &  65.0014 & -54.9378 &0.0047& $\sigma$ & $6.92 \pm 0.10$ & \text{\citet{2002ApJ...579..530W}} & EPN\_S003, etc. \\
87GB 164240.2+262427 & Sy1 &  251.1775 & 26.3202 &0.1441& SE & $7.15 \pm 0.10$ & \text{\citet{2015A&A...575A..13F}} & EPN\_S003 \\
2MASS J16270432+1421249 & Sy1 &  246.7680 & 14.3569 &0.1491& SE & $7.89 \pm 0.04$ & \text{\citet{2022ApJS..263...42W}} & EPN\_S003 \\
2MASS J15394150+5042556 & Sy1 &  234.9232 & 50.7154 &0.2029& SE & $7.86 \pm 0.24$ & \text{\citet{2022ApJS..263...42W}} & EPN\_S003 \\
LEDA 45913 & Sy1 &  198.2745 & -11.1285 &0.0346& DM & $6.48 \pm 0.21$ & \text{\citet{2018ApJ...866...75W}} & EPN\_S003 \\
Z 291$-$51 & Sy1 &  169.7404 & 58.0566 &0.0278& RM & $6.99 \pm 0.10$ & \text{\citet{2014MNRAS.445.3073P}} & EPN\_S003 \\
CSO 498 & Sy1 &  220.7608 & 40.7569 &0.2462& SE & $8.06 \pm 0.24$ & \text{\citet{2022ApJS..263...42W}} & EPN\_S003 \\
LEDA 2816425 & Sy1 &  71.1197 & 12.3532 &0.0899& SE & $7.51 \pm 0.45$ & \text{\citet{2022ApJS..263...42W}} & EPN\_S003, etc. \\
Mrk 142 & Sy1 &  156.3803 & 51.6763 &0.0446& SE & $6.77 \pm 0.10$ & \text{\citet{2005ApJ...625...78H}} & EPN\_S003 \\
ESO 33$-$2 & Sy2 &  73.9940 & -75.5411 &0.0184& $\sigma$ & $7.50 \pm 0.40$ & \text{\citet{2021MNRAS.506.1557W}} & EPN\_S003 \\
IRAS 11119+3257 & Sy1 &  168.6620 & 32.6926 &0.1876& SE & $7.92 \pm 0.36$ & \text{\citet{2022ApJS..263...42W}} & EPN\_S003 \\
2MASS J08525922+0313207 & Sy1 &  133.2468 & 3.2224 &0.2968& SE & $8.41 \pm 0.05$ & \text{\citet{2022ApJS..263...42W}} & EPN\_S003 \\
\end{longtable*}

}

\renewcommand{\thetable}{B\arabic{table}}
\setcounter{table}{0}
\section{Correlations with {Eddington Ratio}}\label{appendixB}

Assuming that the Eddington ratio is a good approximation of the accretion rate, we have $\dot{m}=\lambda_{\rm Edd}=L_{\rm bol}/L_{\rm Edd}$, where $L_{\rm bol}=K_{\rm X}L_{\rm X}$, with the {bolometric correction factor $K_{\rm X}$} derived from \citet{2020A&A...636A..73D}.
We add it as an additional parameter, the relation {changes} to $\log(\tau_{\rm rest}/{\rm s}) = F\log\dot{m} + A\log(M_{\rm BH}/M_\odot) + B\Gamma + C\log(L_{\rm X}/(\rm erg\ s^{-1}))+ D$. 
The results are presented in Table~\ref{tabB}. 
The timescale is found to be anti-correlated with the accretion rate, suggesting that a higher accretion rate leads to more rapid stochastic variability. 
However, none of the BIC values is smaller than the lowest value reported in Table~\ref{tab1}, as this measurement is not independent and its information is largely subsumed by the black hole mass and luminosity. Nevertheless, the result does imply that the coronal structure (traced by $\Gamma$) may exert a more direct influence on the variability timescale than the accretion rate.

\begin{table*}[t]
    \setlength{\tabcolsep}{4mm}
    \centering
    \begin{tabular}{cccccccc}
         \hline
         \hline
         {Model}&$F$&$A$&$B$&$C$&$D$&$\sigma_\epsilon$&BIC  \\
         \hline
1 & $-0.61^{+0.14}_{-0.14}$ & --- & --- & --- & $4.00^{+0.11}_{-0.10}$ & $0.57^{+0.02}_{-0.02}$ & $17.7$ \\
2 & $-0.20^{+0.07}_{-0.06}$ & $1.34^{+0.11}_{-0.10}$ & --- & --- & $-4.98^{+0.70}_{-0.75}$ & $0.39^{+0.02}_{-0.02}$ & $-204.5$ \\
3 & $-0.26^{+0.13}_{-0.13}$ & --- & $-0.67^{+0.10}_{-0.11}$ & --- & $5.54^{+0.26}_{-0.26}$ & $0.54^{+0.02}_{-0.02}$ & $-13.0$ \\
4 & $-1.57^{+0.13}_{-0.15}$ & --- & --- & $1.38^{+0.11}_{-0.10}$ & $-56.10^{+4.44}_{-4.87}$ & $0.40^{+0.02}_{-0.02}$ & $-204.8$ \\
5 & $-0.06^{+0.08}_{-0.08}$ & $1.26^{+0.11}_{-0.11}$ & $-0.20^{+0.08}_{-0.09}$ & --- & $-3.92^{+0.86}_{-0.81}$ & $0.39^{+0.02}_{-0.02}$ & $-204.2$ \\
6 & $-1.33^{+0.15}_{-0.17}$ & $-0.20^{+0.09}_{-0.08}$ & --- & $1.28^{+0.12}_{-0.11}$ & $-51.19^{+4.71}_{-5.16}$ & $0.40^{+0.02}_{-0.02}$ & $-204.8$ \\
7 & $-1.09^{+0.96}_{-1.03}$ & --- & $0.47^{+0.93}_{-1.02}$ & $0.89^{+1.05}_{-0.96}$ & $-37.97^{+35.67}_{-39.18}$ & $0.40^{+0.02}_{-0.02}$ & $-197.4$ \\
8 & $-0.93^{+1.00}_{-1.08}$ & $0.42^{+0.97}_{-1.06}$ & $-0.19^{+0.09}_{-0.09}$ & $0.86^{+1.10}_{-1.00}$ & $-35.89^{+37.35}_{-40.78}$ & $0.39^{+0.02}_{-0.02}$ & $-198.5$ \\

    \hline
    \hline
    \end{tabular}
    \caption{Results from BADDAT regressions with different combinations of variables. 
    The fitted relation is 
    $\log(\tau_{\rm rest}/{\rm s}) = F\log\dot{m} + A\log(M_{\rm BH}/M_\odot) + B\Gamma + C\log(L_{\rm X}/(\rm erg\ s^{-1}))+ D$.
    Each row corresponds to a specific model configuration. 
    {Entries marked as ``--'' indicate that the corresponding variable is not included in the regression.}
    The BIC values are used to assess the model performance.
    }

    \phantomsection
    
    \label{tabB}
    
\end{table*}

\end{document}